\newcommand{\um}{\si{\micro \metre}}
\newcommand{\us}{\si{\micro \second}}
\newcommand{\timePlotWidth}{0.7}

\documentclass{JINST_mod}

\usepackage[hang]{subfigure} 
\usepackage{xspace} 
\usepackage{amsmath} 
\usepackage[section]{placeins} 
\usepackage{siunitx}
\usepackage{arydshln}
\usepackage{enumitem}

\title{Long-term study of backgrounds in the {DRIFT-II} directional dark matter experiment}

\author{
J.~Brack$^a$, E.~Daw$^b$, A.~Dorofeev$^a$, A.C.~Ezeribe$^b$, J.R.~Fox$^c$, J.-L.~Gauvreau$^c$, M.~Gold$^d$, L.~Harmon$^c$, J.L.~Harton$^a$, R.~Lafler$^d$, J.M.~Landers$^c$, R.~Lauer$^d$, E.R.~Lee$^d$, D.~Loomba$^d$, J.A.J.~Matthews$^d$, E.H.~Miller$^d$, A. Monte$^c$, A.StJ.~Murphy$^e$, S.M.~Paling$^f$, N.~Phan$^d$, M.~Pipe$^b$\thanks{Corresponding
author.}~, M.~Robinson$^b$,  S.~Sadler$^b$, A.~Scarff$^b$, D.P.~Snowden-Ifft$^c$, N.J.C.~Spooner$^b$, S. Telfer$^b$, D.~Walker$^b$ and L.~Yuriev$^b$\\
\llap{$^a$}Department of Physics, Colorado State University, Fort Collins, CO 80523-1875, USA\\
\llap{$^b$}Department of Physics and Astronomy, University of Sheffield, S3 7RH, UK\\
\llap{$^c$}Department of Physics, Occidental College, Los Angeles, CA 90041, USA\\
\llap{$^d$}Department of Physics and Astronomy, University of New Mexico, NM 87131, USA\\
\llap{$^e$}SUPA, School of Physics and Astronomy, University of Edinburgh, EH9 3JZ, UK\\
\llap{$^f$}STFC Boulby Underground Science Facility, Boulby Mine, Loftus, Saltburn-by-the-Sea, Cleveland, TS13 4UZ, UK\\
E-mail: \email{m.pipe@sheffield.ac.uk, mark.pipe@physics.ox.ac.uk}}

\abstract{
Low-pressure gas Time Projection Chambers being developed for directional dark matter searches offer a technology with strong particle identification capability combined with the potential to produce a definitive detection of Galactic Weakly Interacting Massive Particle (WIMP) dark matter.  A source of events able to mimic genuine WIMP-induced nuclear recoil tracks arises in such experiments from the decay of radon gas inside the vacuum vessel. The recoils that result from associated daughter nuclei are termed Radon Progeny Recoils (RPRs). We present here experimental data from a long-term study using the DRIFT-II directional dark matter experiment at the Boulby Underground Laboratory of the RPRs, and other backgrounds that are revealed by relaxing the normal cuts that are applied to WIMP search data. By detailed examination of event classes in both spatial and time coordinates using $5.5$~years of data, we demonstrate the ability to determine the origin of 4 specific background populations and describe development of new technology and mitigation strategies to suppress them.  
}

\keywords{WIMPs; dark matter; TPC}

\begin{document}

\section{Introduction, DRIFT and DRIFT analysis}
\label{sec:introduction}
Determination of the nature of particle dark matter represents one of the major challenges in particle physics and cosmology. For success it is increasingly recognised that direct searches for the most favoured class of candidate, Weakly Interacting Massive Particles (WIMPs), should include information that can demonstrate a Galactic origin of observed candidate events~\cite{ahlen2010}.  The motivation for this has been highlighted recently by the observation, in several experiments without such capability, of unexpected backgrounds or signals~\cite{dm2012, aalseth2011, cdms}. Technologies able to determine the direction of low-energy nuclear recoils in a target provide a route towards such a Galactic signature and hence perhaps towards definitive evidence for WIMPs~\cite{morgan2005}.

The DRIFT (Directional Recoil Identification from Tracks) experiments at the Boulby Underground Laboratory have, since 2001, pioneered development of low-pressure gas Time Projection Chambers (TPCs) to achieve this directionality~\cite{alner2004, alner2005, burgos2009a, burgos2009b, daw2012}.  Activity has also expanded worldwide to include at least 4 other directional WIMP search efforts (DMTPC, MIMAC, NEWAGE, D3)~\cite{ahlen2010,cygnus2011}.   These devices all have strong particle identification capability, and this opens the possibility of examining the origin and nature of rare nuclear recoil-like backgrounds to the dark matter search, particularly from radon contamination and most significantly from a class of events called radon progeny recoils (RPRs). Preliminary observation and characterisation of RPRs in DRIFT is described in Burgos et al. (2007) \cite{burgos2007}. Here we present more detailed work on backgrounds in DRIFT-II, focussed on long-term analysis over $5.5$~years in relation to technology changes made to the detector.  

\begin{figure}[ht]
\begin{center}
\includegraphics[width=0.47\textwidth]{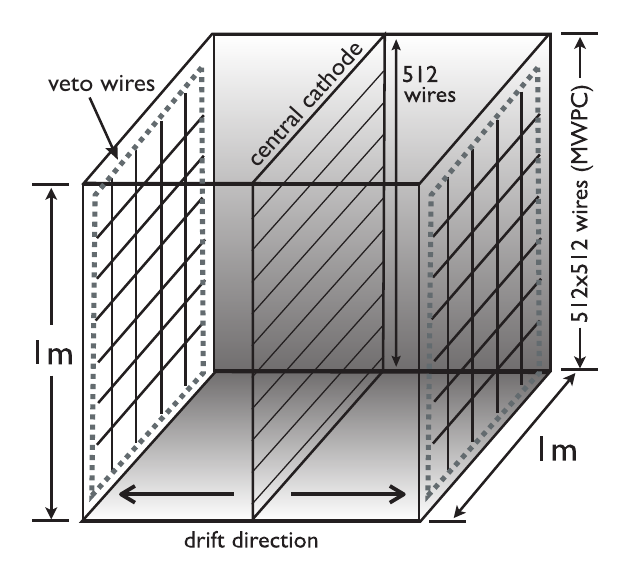}
\includegraphics[width=0.37\textwidth]{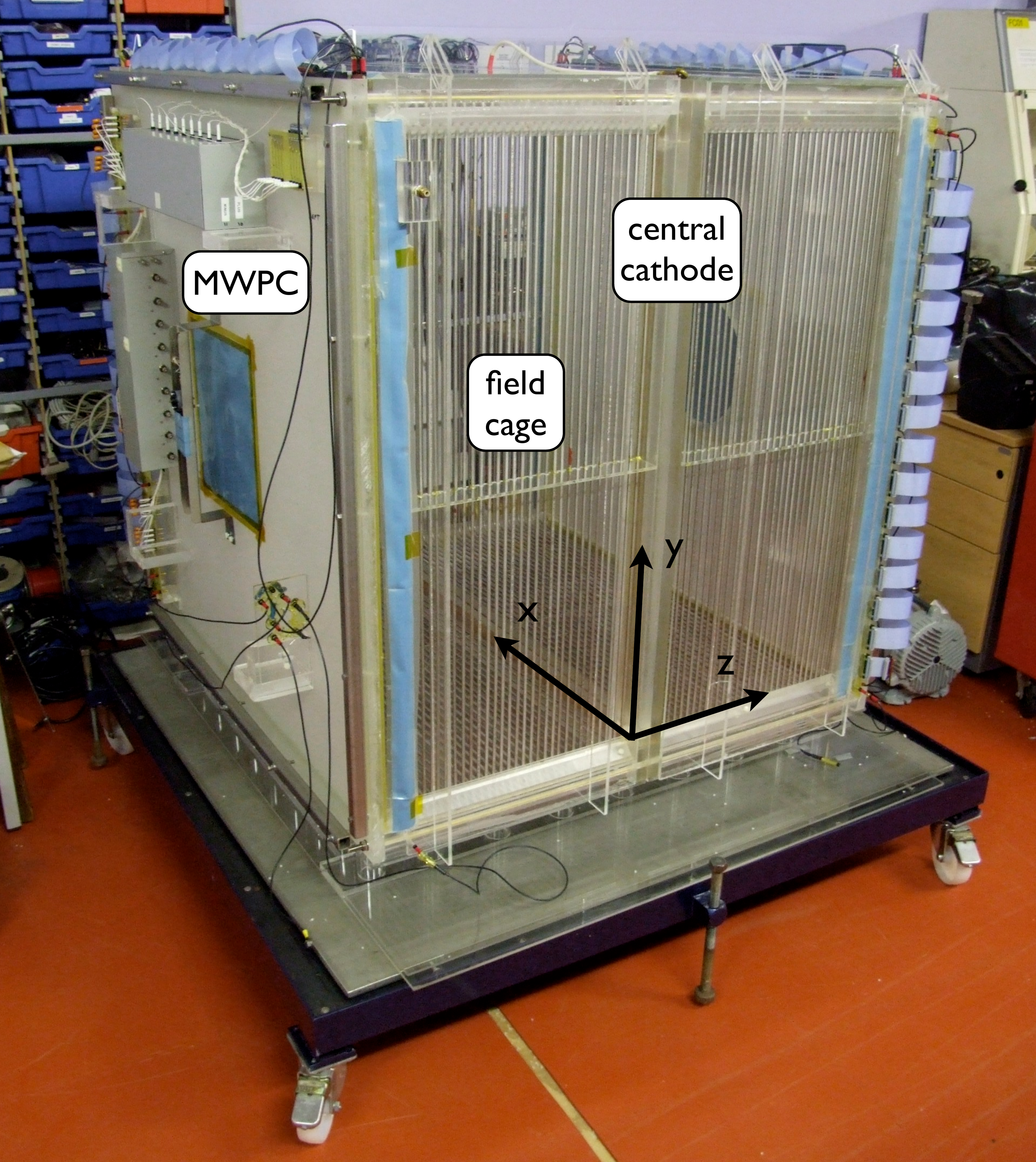}
\caption{The left schematic illustrates the geometry of the 1\,m$^3$ DRIFT-II detector. The photo on the right shows the DRIFT-II module removed from the steel vacuum vessel. Further details are discussed in the text.}
\label{fig:schematic}
\end{center}
\end{figure}

Figure~\ref{fig:schematic} shows the DRIFT-II detector as originally installed in Boulby at $2805$~m.w.e. depth.  Full technical details are given in Alner et al. (2005)~\cite{alner2005}.  Briefly, DRIFT-II comprises a 1.5~m by 1.5~m by 1.5~m stainless steel vessel containing two 1.0~m by 1.0~m back-to-back Multi-Wire Proportional Chambers (MWPCs) each 50\,cm from a central 1.0~m by 1.0~m cathode plane contained within a field-cage and supplied with -28.4\,kV.  Each MWPC has 3 layers of 512, 2\,mm-spaced stainless steel wires - a central anode plane with wires having 20\,\um~diameter and two perpendicular grid planes of 512 100\,\um~wires; the inner grid plane faces into the main detector volume, the outer plane faces away. The anode-grid separation is 10\,mm on either side, with $2.84$~kV applied to the grid planes to provide avalanche gain of $\sim 1000$.  

DRIFT is a negative ion TPC (NITPC) normally operated at a pressure of 40~Torr. Electronegative CS$_2$ gas captures electrons liberated in particle interactions in the fiducial volume, acting as the charge carrier and reducing diffusion along the drift path to thermal levels \cite{Snowden2013}. CS$_2^-$ anions are then drifted to the nearest MWPC, where the release of the electrons results in avalanche events recorded in two dimensions on the orthogonal grid and anode readout wires. Track extent in the third (z) dimension is calculated from the pulse durations on the anode wires according to
\begin{equation}
\Delta z = \frac{\mu E \Delta t}{P} \text{,}
\label{eqn:z}
\end{equation}
where $\Delta t$ is pulse duration, $\mu$ is the reduced mobility, $E$ is the electric field ($550$~V/cm), and $P$ is pressure (here, 40~Torr). In practise, $\Delta z$ also has contributions from diffusion, the shaping electronics, and the curved isochrones that result from the differing drift path lengths for ions arriving directly above, and directly in between, anode wires \cite{Snowden2013}. 52 (41) wires on the edges of the MWPC grid (anode) planes are grouped to provide x-y veto signals, and z-direction vetoing is under development~\cite{Snowden2014}.

\begin{figure}[ht] \begin{center}
\includegraphics[width=0.8\textwidth]{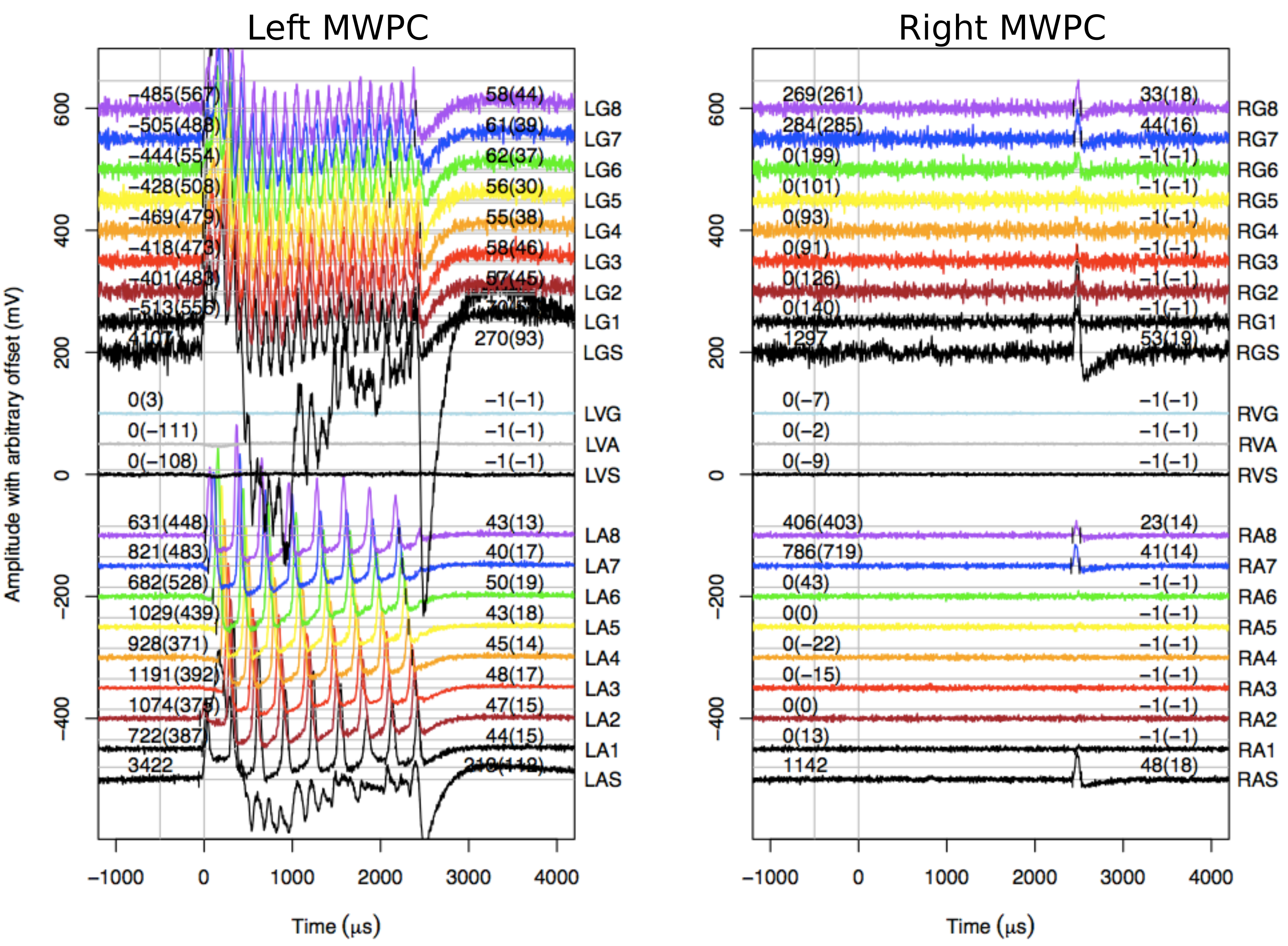}
\caption{A typical alpha-tagged recoil event. The left-hand side MWPC contains a straight alpha track that `wraps around', producing multiple hits on each channel. The right-hand side contains a recoil-like event that passed all the WIMP analysis cuts and hit two channels (RA7 and RA8), giving it a range of ($2 \times 2$~mm wire pitch $- 1$~mm $=$) 3~mm. The coincidence in time between the left and right events at 2500\,\us~confirms these as two parts of the same event with common origin at the z position corresponding to the central cathode (50~cm).}
\label{fig:tagged_rpr_event}
\end{center}
\end{figure}

A description of the DRIFT-II data acquisition is outlined in Daw et al. (2012)~\cite{daw2012}.  Key here is that MWPC wires are grouped such that every eighth wire is summed together. Therefore, eight adjacent wires sample a distance of 16\,mm, which is sufficient to contain all nuclear recoil tracks of interest. Figure~\ref{fig:tagged_rpr_event} shows an event display for a tagged RPR event, which is described in detail in section~\ref{sec:backgroundEventClasses}. It is shown here to illustrate the difference in DRIFT data between the alpha particle on the left, and the nuclear recoil on the right of the detector. The output from the two MWPCs, left and right, can be seen, divided between 8 grid channels (top lines: G1--G8) and 8 anode channels (bottom lines: A1--A8) plus veto signals (centre three lines). Alpha particles such as that appearing on the left have ranges $\sim 300$~mm and `wrap around', producing multiple hits above the fixed $5$~mV analysis threshold on each channel. Typical recoil tracks, such as those from scattering of C, S or F nuclei by fast neutrons or from WIMPs (see the right panel of figure~\ref{fig:tagged_rpr_event}), range a few mm and hit $<8$ channels. The W-value of the 30:10~Torr CS$_2$:CF$_4$ gas mixture is measured to be $25.2\pm0.6$~eV~\cite{Pushkin2009}, which means that a fast-moving $5$~MeV$_{r}$ $\alpha$~particle produces $\sim 200$k~NIPs (number of ion pairs). The nuclear quenching factor reduces the ionisation for slow-moving particles. For example, a $60$~keV$_{r}$ nuclear recoil produces $\sim 1$k~NIPs in the gas \cite{Hitachi2008}.

\section{Event selection and background populations}
\label{sec:event_selection}
\begin{figure}[ht]
\begin{center}
\includegraphics[width=0.5\textwidth]{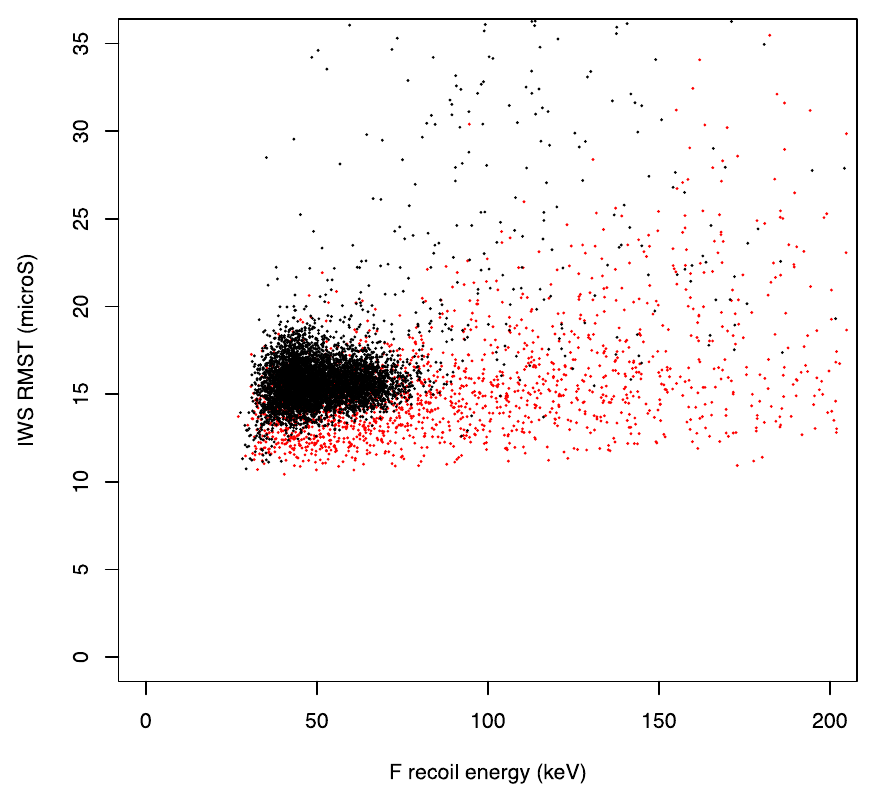}
\caption{RMST vs. fluorine recoil energy distribution for candidate neutron recoil events from a $^{252}$Cf calibration run (red), and events from the $47.4$~day WIMP search run (black) presented in Daw et al. (2012)~\cite{daw2012}. The higher RMST of the events from the WIMP search run is a consequence of these events occurring on the central cathode, and hence suffering maximal diffusion when drifted to the MWPC detector. }
\label{fig:rmst_nips}
\end{center}
\end{figure}

A particularly important parameter for nuclear recoil events is the root mean square in time (RMST) of the pulse, which measures the time spread of charge arriving at the MWPC. This provides some information on an event's $z$-position, because charge originating further from the MWPC has further to drift, and hence suffers greater diffusion, which broadens the pulse in time and gives the event a higher RMST. The distribution of events in the RMST-NIPs plane is effective at separating nuclear recoils in the bulk of the gas (for example, from WIMP or $^{252}$Cf calibration neutron interactions) from background events originating on or near the central cathode. This can be seen in figure~\ref{fig:rmst_nips}, which was generated using the full set of WIMP analysis cuts described in Daw et al. (2012) \cite{daw2012} (see also Snowden-Ifft et al. (2003) \cite{Snowden2003} and Burgos et al. (2009) \cite{burgos2009c}).

The full WIMP analysis cuts used to select events appearing in figure~\ref{fig:rmst_nips} eliminate all background events with low RMST. However, they also reduce the efficiency for accepting calibration neutrons by a factor of at least $2.7$. This number is a lower limit because, with only basic cuts in place (full containment, signals within the range of the digitisers, etc), gamma rays from the $^{252}$Cf neutron source can masquerade as neutron recoils. This precludes the removal of gamma-rejection cuts when calculating the pre-cuts neutron acceptance rate, causing it to be underestimated. Using the full set of WIMP analysis cuts described in Daw et al. (2012) \cite{daw2012}, the removal of background events at low RMST allows the definition of a background-free signal region that is $\sim 20\%$ efficient at accepting neutron calibration recoils passing all other cuts.

\begin{figure}[ht]
\begin{center}
\includegraphics[width=\textwidth]{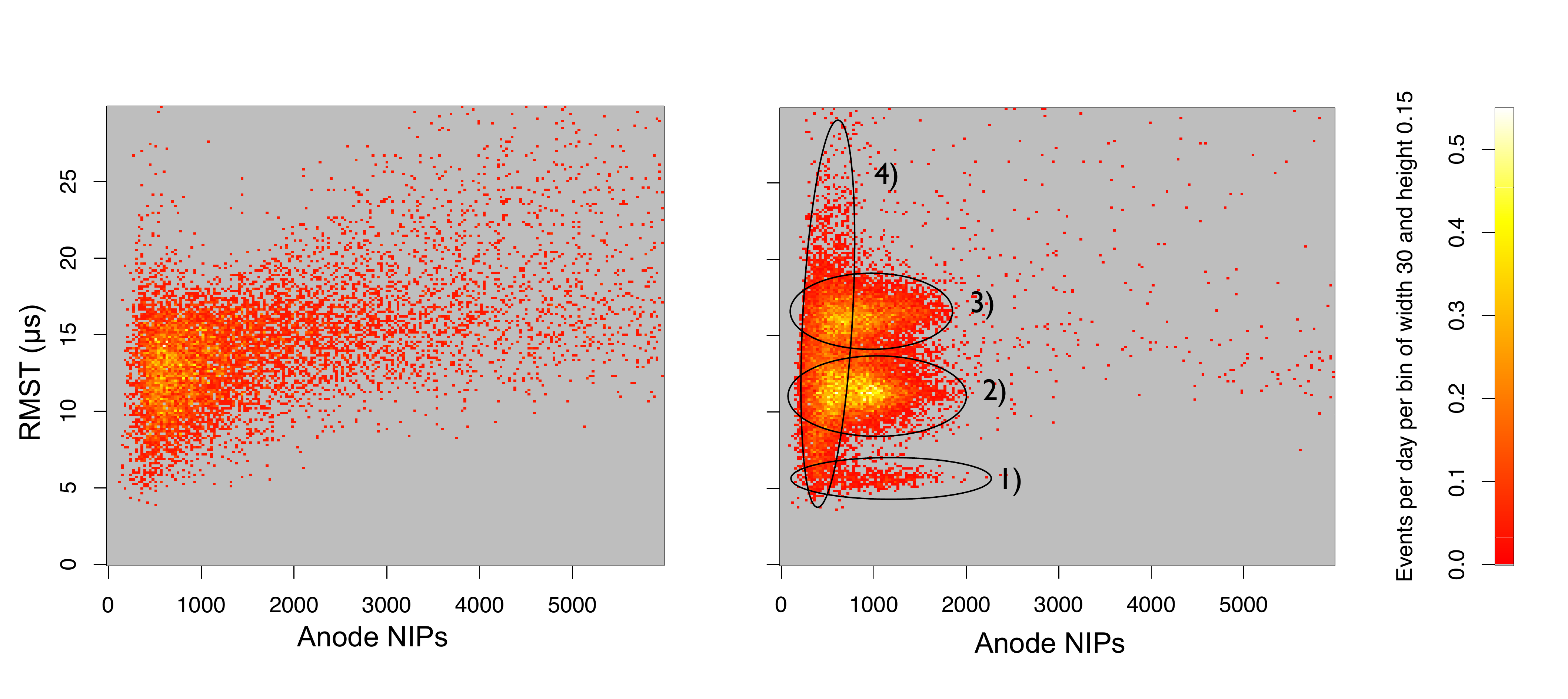}
\caption{Comparison of WIMP run data to neutron data with minimal cuts: (left) from 1.84 days of neutron exposure with events distributed over a wide range of NIPs and RMST, (right) 47.26 days of WIMP run data showing remaining background events in four distinct populations (see text).}
\label{fig:moving_bg_plots_example}
\end{center}
\end{figure}

The reduced neutron/WIMP acceptances caused by the full WIMP analysis cuts motivates the study of background populations that only appear after these cuts have been relaxed. In contrast to the work presented in Daw et al. (2012)~\cite{daw2012}, in this work analysis proceeded using such a reduced set of cuts designed to accept all potential recoil-mimicking events. Specifically, several low-efficiency, high-purity `pulse shape' cuts were removed, which had the effect of  revealing event populations that would be rejected by the full set of WIMP analysis cuts. Figure~\ref{fig:moving_bg_plots_example} shows the results of this analysis using background data taken with shielding erected, and neutron source data used to produce calibration nuclear recoils. 

While the neutron data form a single population, events from the background, with a factor $\sim 12$ lower rate, form four distinct recoil-like populations in the RMST-NIPs plane. There is some overlap between the population~4 and the low-energy events in the other three populations, which prevents the calculation of absolute rates for populations 1, 2 and 3. In section~\ref{sec:driftOperation} we present a largely qualitative analysis of the change in rates of the four populations over a period of $\sim 4$~years, during which time modifications were made to the MWPCs and central cathode. A threshold of $700$~NIPs ($\sim 50$~keV$_{\mathrm{r}}$) was implemented when counting events in populations 1, 2 and 3 to reduce the population 4 contamination. As figure~\ref{fig:moving_bg_plots_example} shows, this is comfortably above the drop in efficiency at low energy caused by the $20$~mV hardware trigger. By examining the change in the rates of the four populations caused by each of the detector modifications, and comparing these to the rates of well-understood alpha event classes detailed in section~\ref{sec:backgroundEventClasses}, we seek to explain the origin of the four populations. The following summarises the state of knowledge about them prior to this work.

\begin{figure}[ht]
\vspace{-0cm}
\begin{center}
\includegraphics[width=0.45\textwidth]{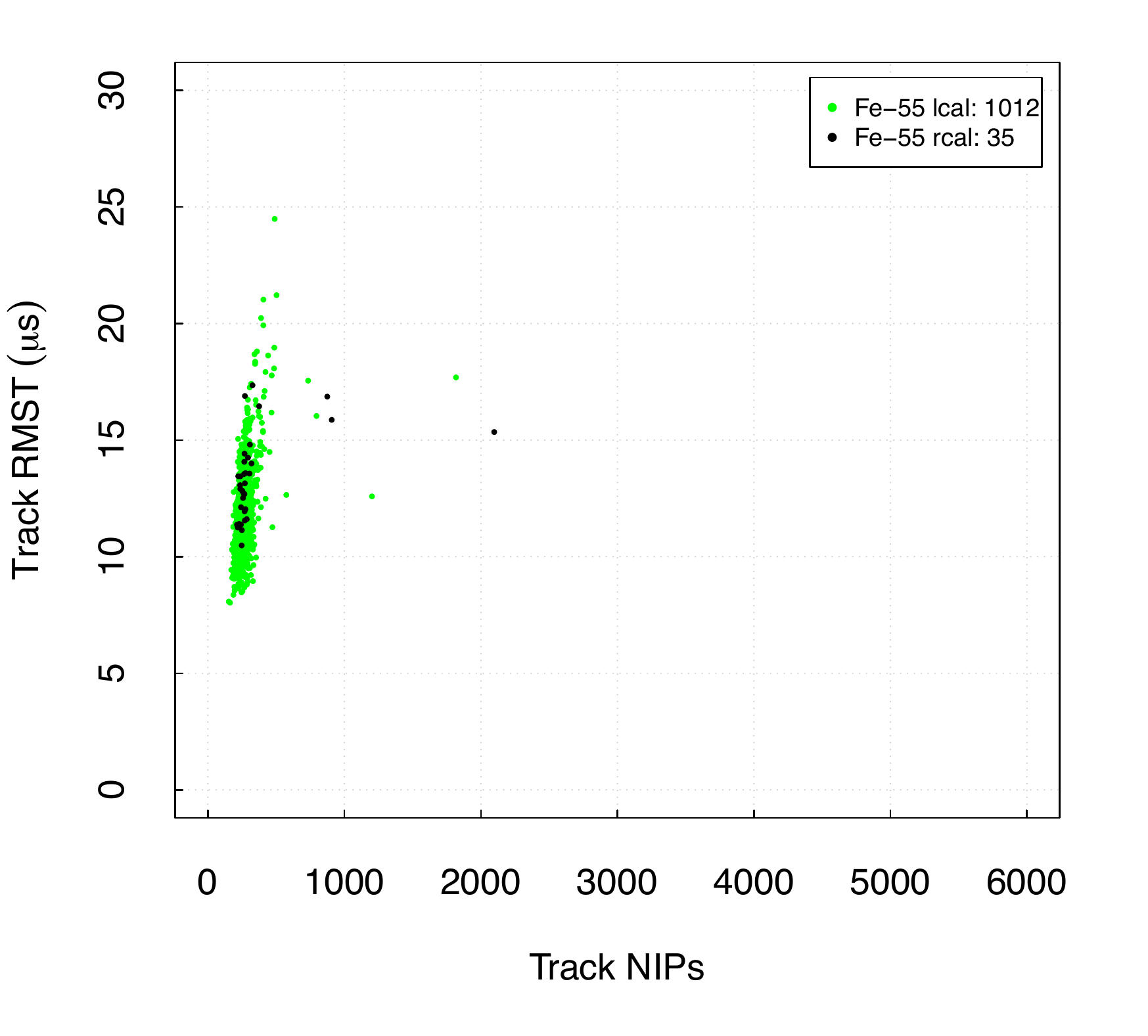} 
\caption{Analysis of $^{55}$Fe gamma rays on the left MWPC with the reduced set of cuts designed to select recoil-like events. Light grey (green online) points from 40 mins exposure (rate of 0.42$\pm$0.01\,Hz) with distribution consistent with the population 4 events. Black points are from 40~mins of background in untriggered mode (rate of 0.014$\pm$0.002\,Hz observed).}
\label{fig:wimp_analysis_of_fe55}
\end{center}
\end{figure}

{\bf Population 1} --- These events have an RMST of $\sim$5\,\us, and are consistent with noise due to internal detector discharges with pulse shapes consistent with the characteristic fast time of the shaping amplifiers. In the normal WIMP analysis, these events are all clearly separable with minimal loss to neutron efficiency. {\bf Population 2} --- Events here are found to have different pulse shape characteristics from neutron-induced recoils in the same NIPs vs. RMST region. Although this difference can be used to eliminate this population in WIMP search data~\cite{daw2012}, the cuts required reduce the recoil (and WIMP) detection efficiency by a factor $> 2.7$. Improving understanding of population 2 is therefore of particular interest. {\bf Population 3} --- Events here are indistinguishable from neutron-induced recoils in the same region of NIPs vs. RMST implying they are true nuclear recoils, likely from decays on the central cathode (RPRs). The signal region in the NIPs vs. RMST plane required to exclude these events significantly reduces the recoil detection efficiency~\cite{daw2012}. {\bf Population 4} --- These are consistent with low-energy ($\lesssim50$~keV) electrons from gamma backgrounds. This was confirmed using 40~minutes of exposure to an internal $^{55}$Fe (5.9\,keV x-ray) source using the reduced set of cuts (see figure~\ref{fig:wimp_analysis_of_fe55}).  A cluster of events above background is seen with distribution indeed consistent with population 4.

\section{Alpha event classes in DRIFT-II}
\label{sec:backgroundEventClasses}
\begin{figure}[ht]
\begin{center}
\includegraphics[width=0.65\textwidth]{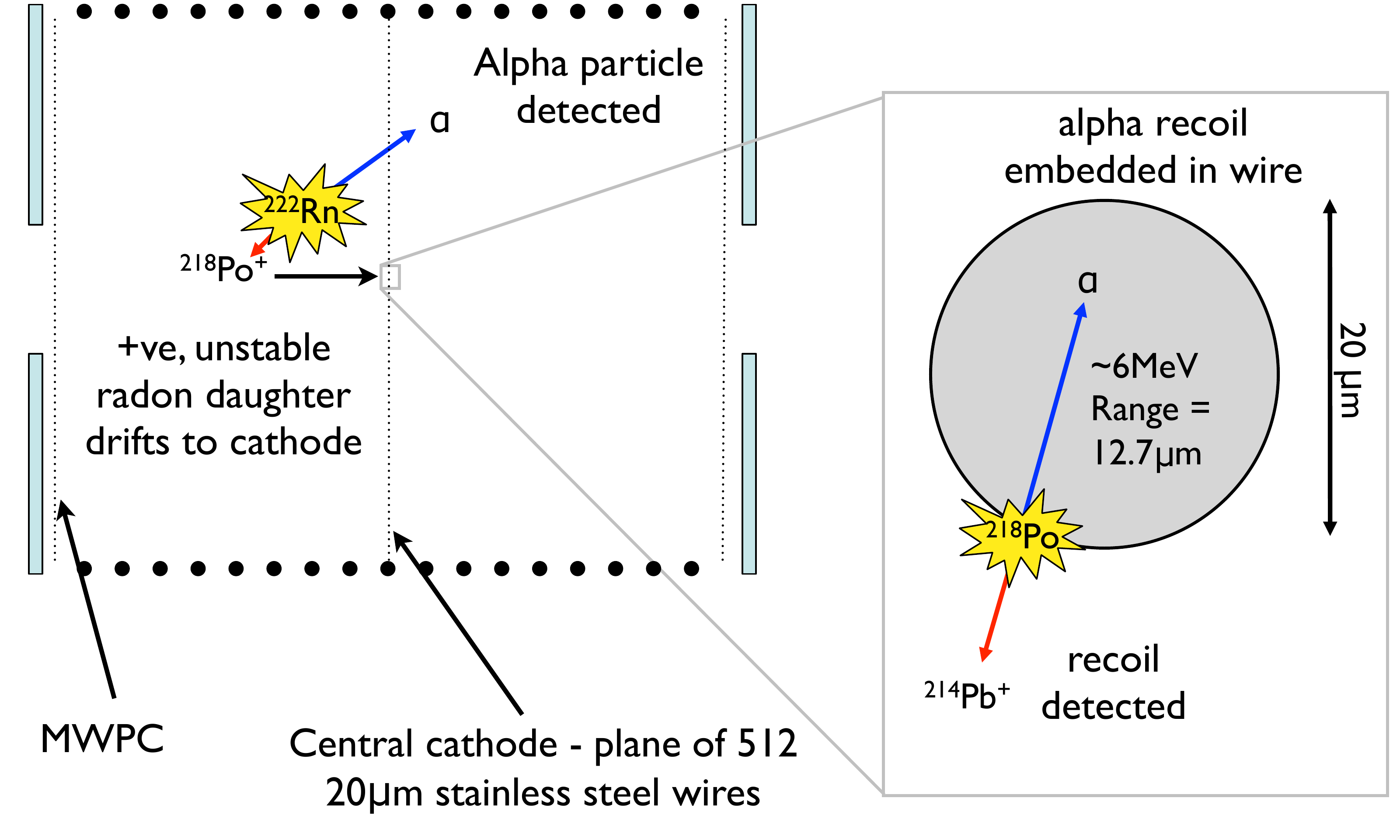}
\caption{Schematic of RPR background in DRIFT-II. Rn decay can result in a positively charged ion that drifts to, and plates out on, the 20\,\um~stainless steel central cathode wires. The daughter then alpha decays.  If the alpha is oriented into the wire, a WIMP-like nuclear recoil occurs without the tell-tale alpha event.}
\label{fig:rpr_schematic} 
\end{center}
\end{figure}

The RPR mechanism presented in Burgos et al. (2007)~\cite{burgos2007} attributes the black points in figure~\ref{fig:rmst_nips} to the decay of positively-charged radon daughter radionuclides on the central cathode. Figure~\ref{fig:rpr_schematic} illustrates this process. $^{222}$Rn, suspended in the gas, decays via the emission of a 5.6~MeV alpha particle. Although this can reliably be identified, the daughter is an unstable $^{218}$Po atom that can be produced positively charged~\cite{burgos2007}. In this case the $^{218}$Po$^+$ ion is attracted to, and plates out on, the negatively charged central cathode wires, where it decays with a half-life of 3.1~minutes by emission of a 6.11\,MeV alpha particle. Simulations with the Monte Carlo simulation code Stopping Range of Ions in Matter (SRIM 2010) show that an alpha particle of this energy has a range of $\sim$12.7\,\um~in the $20$~\si{\micro \metre} diameter stainless steel central cathode wires \cite{srim}. This geometry, illustrated in figure~\ref{fig:rpr_schematic}, and the random orientation of the alpha decay results in a number of possible event types, which are detailed later in this section.

The RPR background arises in the case where the alpha particle is ejected into the wire and lost to the detector, while the recoiling daughter nucleus propagates into the fiducial volume, typically producing $\sim 1000$~NIPs ($\sim$60\,keV$_{\mathrm{r}}$). Recoils of this energy are potentially indistinguishable from WIMP-induced nuclear recoils~\cite{burgos2007}. Conversely, if the alpha particle is emitted into the gas, then it is easily identified and can be used to `tag' the event as an RPR and reject it.

\begin{figure}[bth]
\begin{center}
\subfigure{ \includegraphics[width=0.45\textwidth]{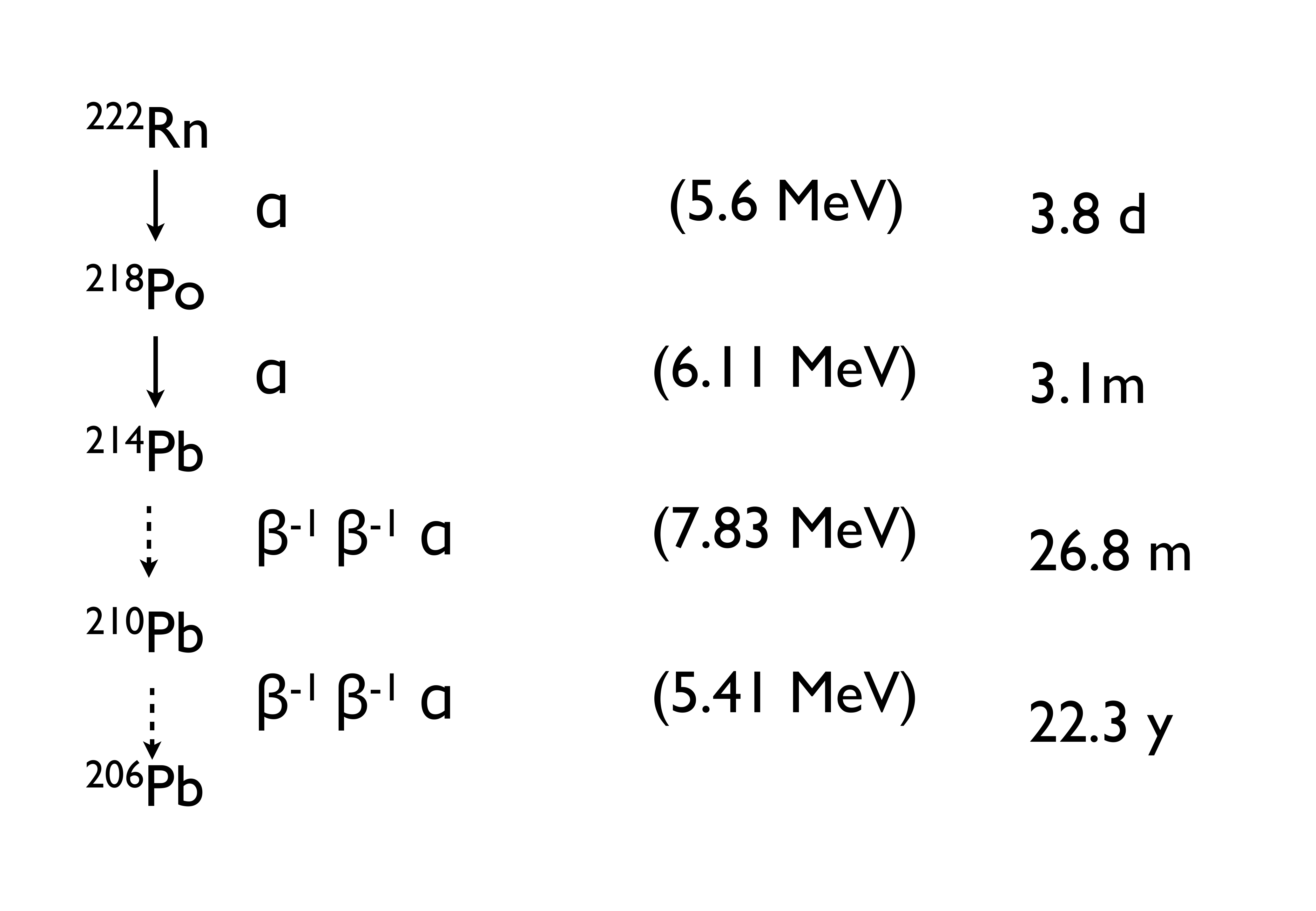} }
\subfigure{ \includegraphics[width=0.45\textwidth]{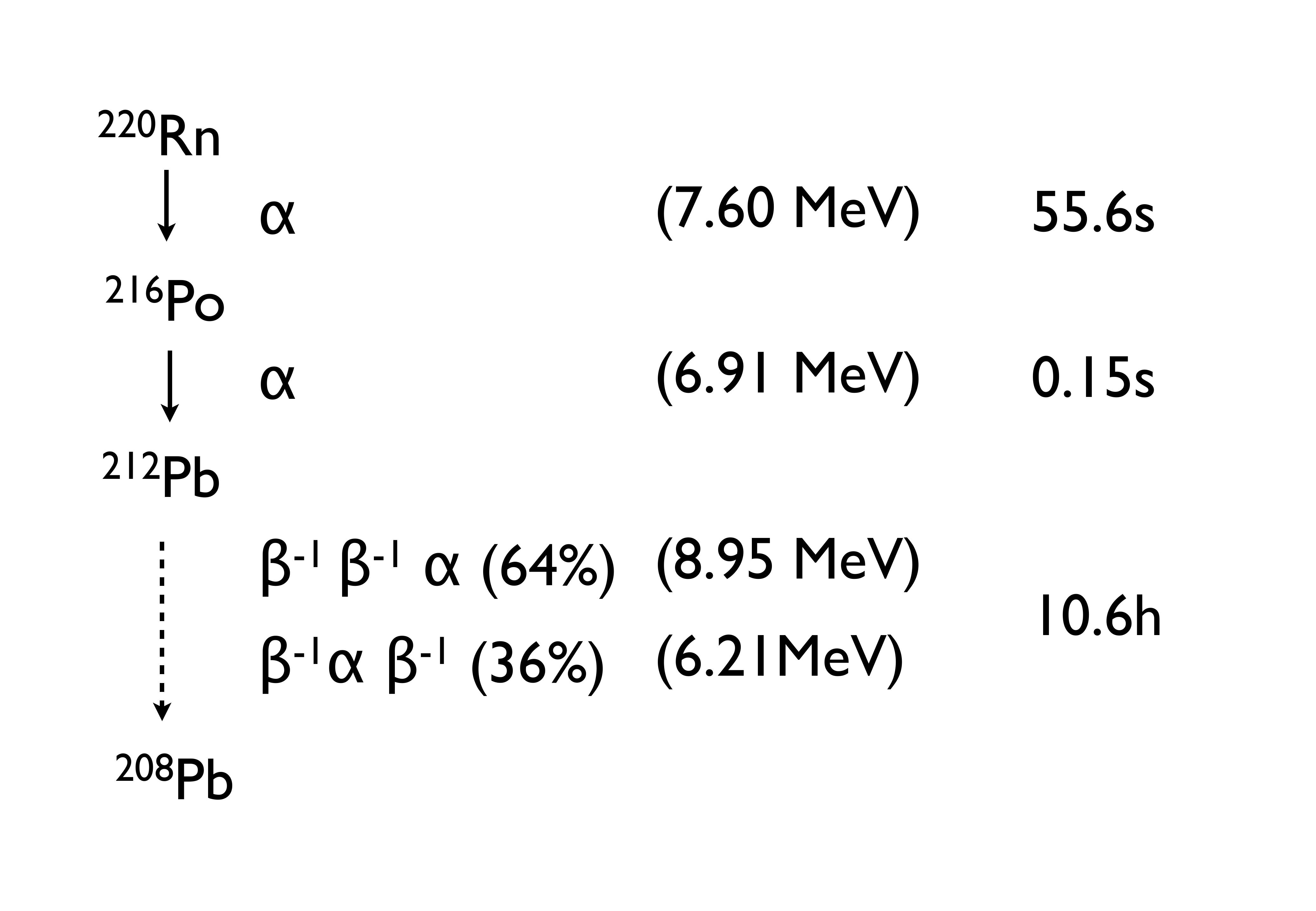} }
\caption{U ($^{222}$Rn) and Th ($^{220}$Rn) decays relevant to DRIFT. $^{222}$Rn and its daughters produce four alpha particles with energies of 5.4 to 7.8\,MeV. $^{220}$Rn produces three alpha particles of 6.2 to 9.0\,MeV. \cite{NSR2014}.}
\label{fig:radon_decay_chains}
\end{center}
\end{figure}

Detector components are known to emanate radon into the vacuum vessel~\cite{sadler2014}. As shown in figure~\ref{fig:radon_decay_chains}, the $^{222}$Rn decay chain results in further radioisotopes that may be produced positively charged and hence plate out on the central cathode wires \cite{Hopke1996}. In the end the decay chain leads to $^{206}$Pb via $^{210}$Pb, which decays to $^{206}$Pb by emission of a 5.41\,MeV alpha particle with a half-life of 22.3~years. This process may therefore result in long-term accumulation of $^{210}$Pb on the central cathode wires, although the $^{210}$Pb on the wires is thought to be dominated by a component leftover from the manufacturing process. Based on this, two main classes of Rn background are predicted: a prompt RPR background of WIMP-like recoils within minutes or hours of the decay of the Rn in the gas, and a RPR background that increases over time due to the accumulation of long-lived $^{210}$Pb on the wires. The latter would remain long after Rn is removed from the detector. A strong correlation between the rate of alpha-tagged RPRs and the rate of population 3 events would imply that these events share a common production mechanism, namely the decay of radon daughters on the central cathode.

\begin{figure}[ht]
\vspace{0.5cm}
\begin{center}
\includegraphics[width=0.8\textwidth]{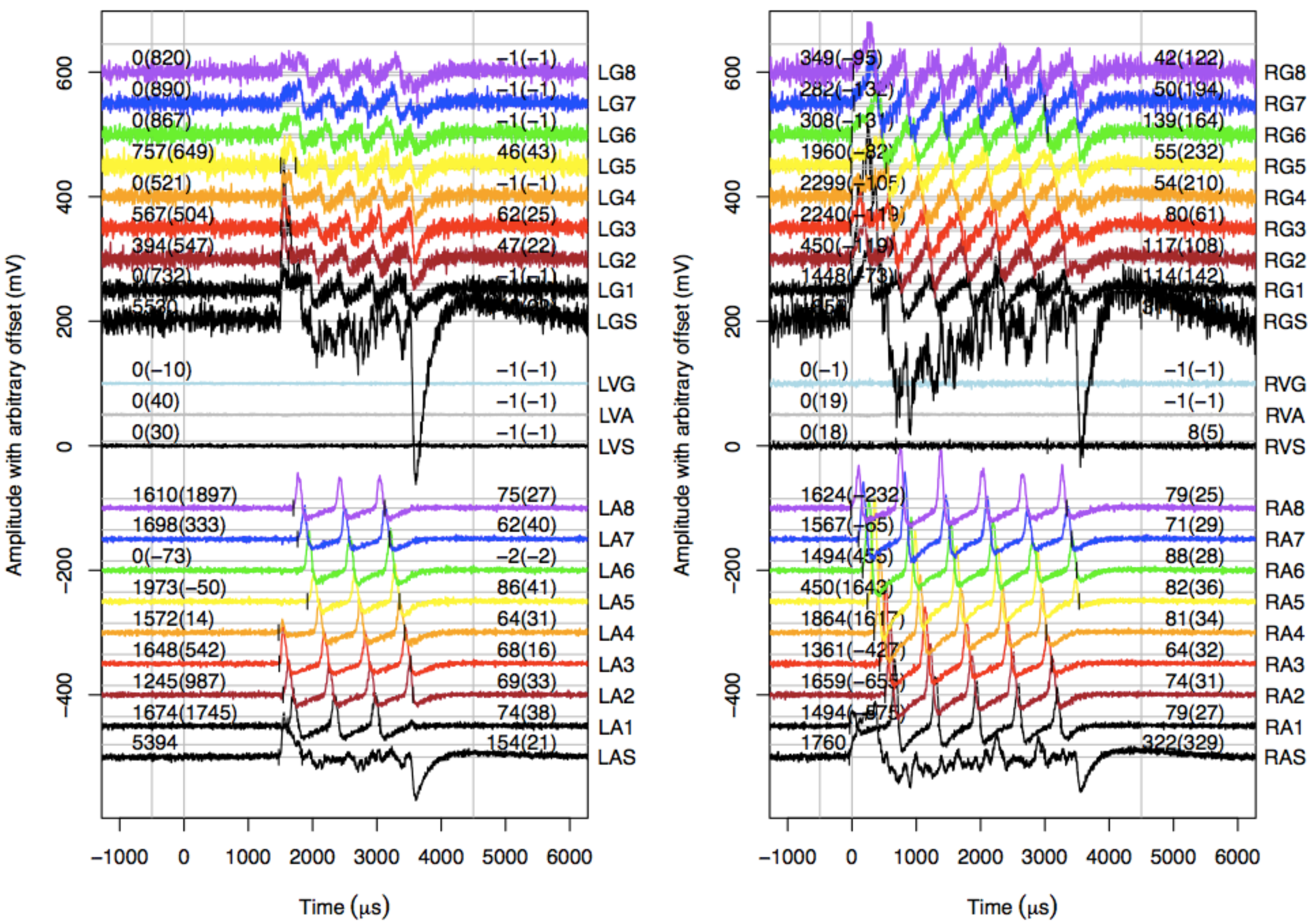}
\caption{A typical GPCC event, in which a distinct alpha signal is seen on both MWPCs. The DAQ is triggered by the first anode channel to cross a fixed threshold of $20$~mV on either side (here, the right), and times here are relative to this trigger time. The track ends simultaneously on each side, at $\sim$3500\,\us, indicating that these are two halves of the same track that join at the central cathode.}
\label{fig:GPCC_event} 
\end{center}
\end{figure}

We consider three alpha event categories as follows, each of which are isolated with a custom-built set of cuts independent of those described here and in Daw et al. (2012) \cite{daw2012}. (1)~{\bf Gold-plated cathode-crossing alphas (GPCCs)} -- these are fully contained within the fiducial volume, but pass through the central cathode (see figure~\ref{fig:GPCC_event}). Such events will result in an alpha track in each MWPC, a signature that, for alpha particles in the energy range of interest ($\approx 5$~MeV), can only be produced by the alpha decay of neutral radionuclides suspended in the target gas, yielding a direct measure of the level of radon contamination in the target gas \cite{burgos2008,sadler2014}. The GPCC selection requires all $8$~channels to be hit on both sides, and late-time coincidence of the track ends (to within $150$~\si{\micro \second}, or $0.8$~cm in z).  (2)~{\bf Alpha-tagged recoil events} -- these are from radionuclides deposited on the central cathode of the detector, in particular where a decay on the surface of a central cathode wire results in an alpha being emitted into one side of the fiducial volume, whilst the daughter nucleus recoils into the other side (see figure~\ref{fig:tagged_rpr_event}), yielding an associated alpha and nuclear recoil (RPR) pair. This type of event can only be created by the decay of an atom on the surface of the central cathode and thus offers a direct measure of contamination on the central cathode wires. Alpha-tagged recoils are selected as having all channels hit on one side (the alpha track), and an event passing the full set of recoil cuts in late-time coincidence (again, within $150$~\si{\micro \second}, or $0.8$~cm, in z) on the opposite side. (3)~{\bf 1-MWPC alpha events} -- these are alpha particles fully contained within a single side of the fiducial volume, defined as having no hits on the veto channels, one side with all 8 channels hit, and no hits on the opposite side.  They must originate inside the fiducial volume from radioactive decays in the target gas, or on the surface of the electrodes (MWPC or central cathode) that define the fiducial volume. The left-hand panel of figure~\ref{fig:tagged_rpr_event} depicts a typical MWPC alpha event.

\section{DRIFT-II detector operations and results of long-term background studies}
\label{sec:driftOperation}
\begin{figure}[ht]
\begin{center}
\includegraphics[width=\timePlotWidth\textwidth]{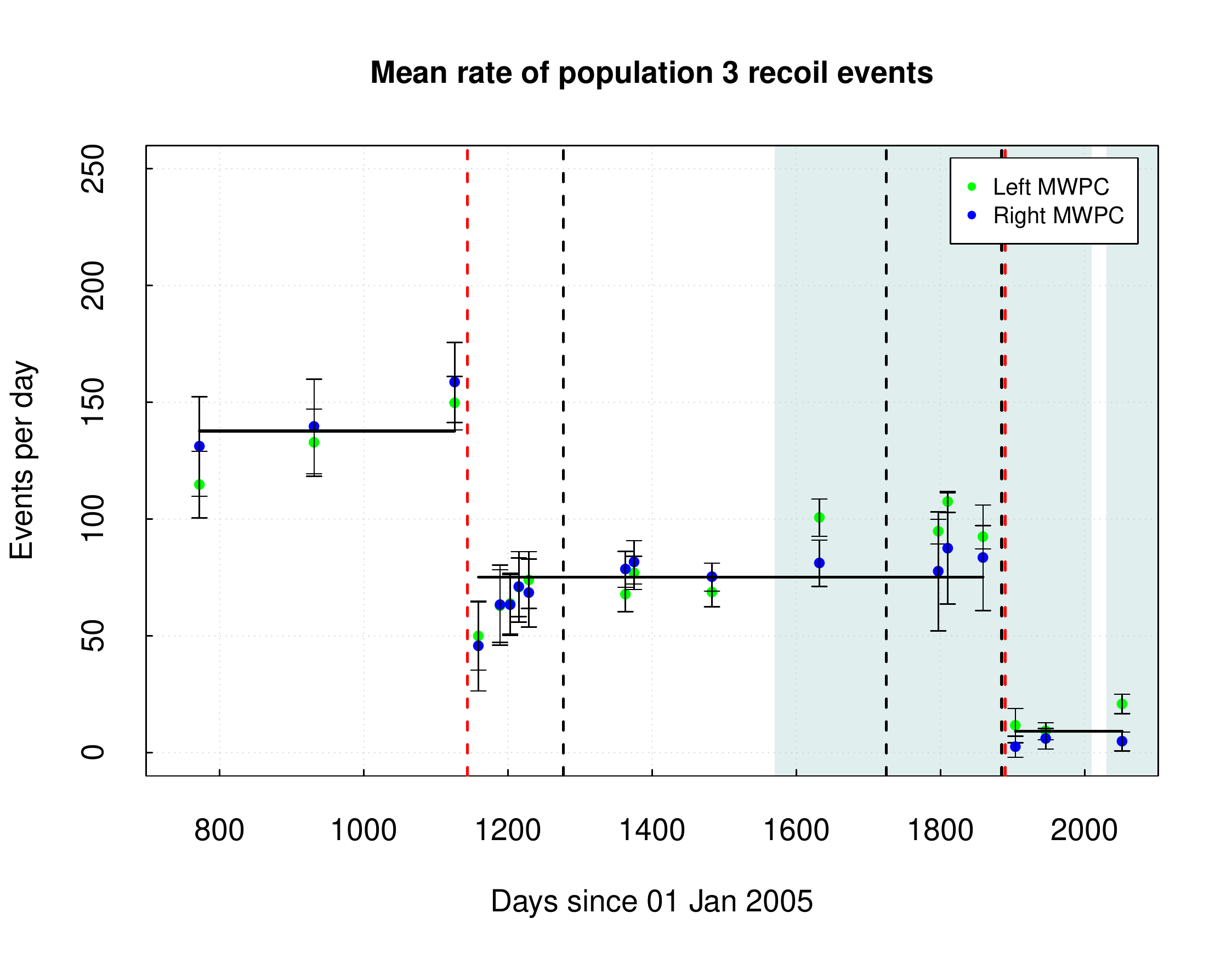}
\caption[Mean event rate of population 3 recoil events]{Population 3 event rate with mean (black horizontal lines) of $138\pm6$~\si{events\per \day} with the wire central cathode prior to etching, $75\pm3$~\si{events\per \day} after etching, and $9\pm3$~\si{events\per \day} with thin film cathode. Vertical dashed lines indicate described actions. Blue regions are CS$_2$-CF$_4$ data, white regions are pure CS$_2$.
}
\label{fig:mean_event_rate-cathode_events} 
\end{center}
\end{figure}

\begin{figure}[ht] \begin{center}
\includegraphics[width=\timePlotWidth\textwidth]{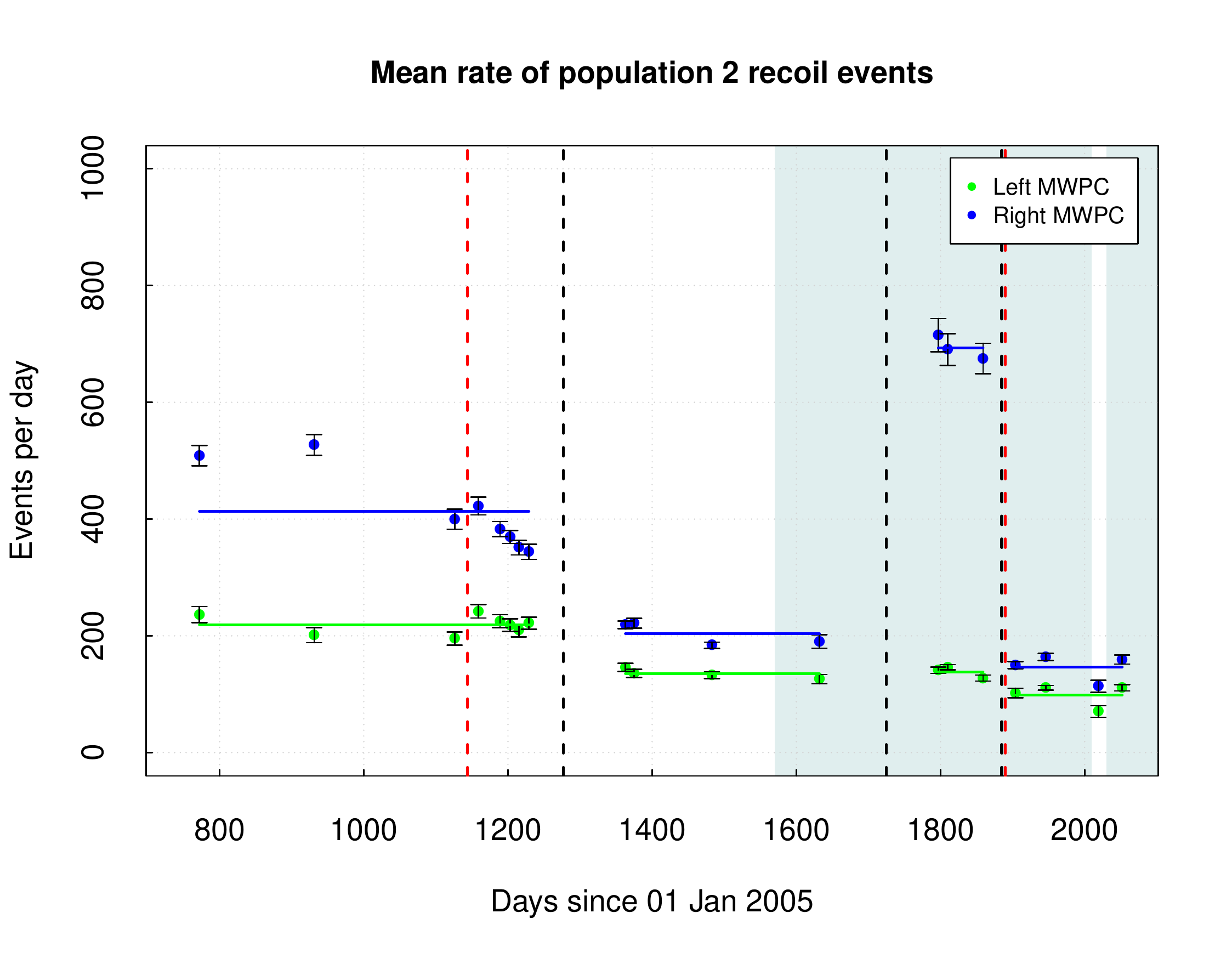}
\caption{Rate of population 2 events with mean on left (right) MWPC of 219$\pm$6 (413$\pm$25) events per day before MWPC etching, 135$\pm$4 (204$\pm$10) events per day after etching, 138$\pm$6 (693$\pm$12) events per day with replacement MWPC on the right and 99$\pm$10 (147$\pm$11) events per day with the original MWPC replaced. Vertical dashed lines indicate actions described in section~3. Blue regions are CS$_2$-CF$_4$ data, white regions are pure CS$_2$.
}
\label{fig:mean_event_rate-mwpc_events} 
\end{center} \end{figure}

\begin{figure}[ht]
\begin{center}
\includegraphics[width=\timePlotWidth\textwidth]{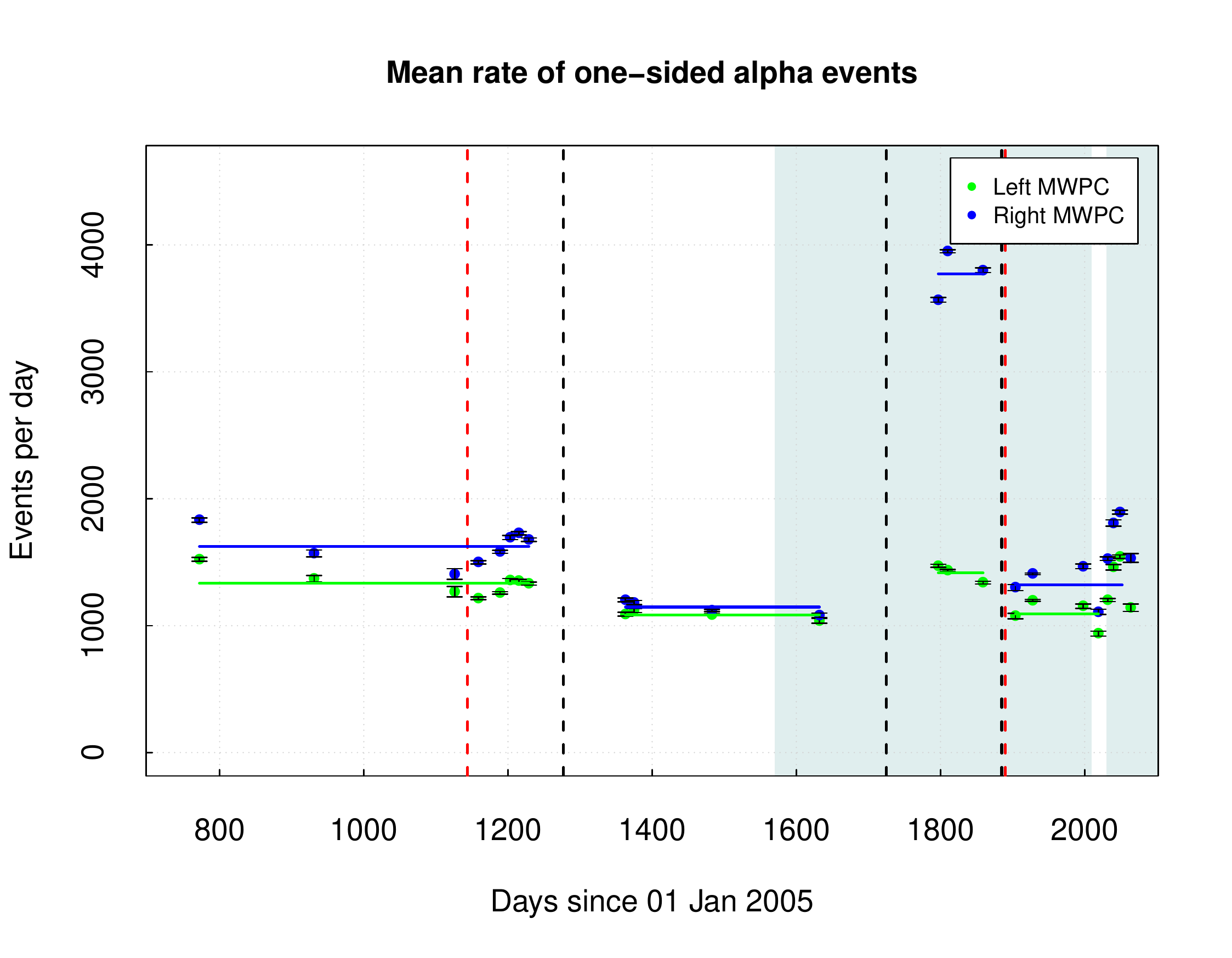}
\caption{Rate of 1-MWPC alpha events with mean on left (right) MWPC of 1334$\pm$33 (1624$\pm$48) events per day before MWPC etching, 1084$\pm$17 (1147$\pm$28) after MWPC etching, 1415$\pm$40 (3773$\pm$111) with replacement MWPC on the right and 1092$\pm$57 (1321$\pm$79) with original MWPC replaced. Vertical dashed lines indicate actions described in section 3. Blue regions are CS$_2$-CF$_4$ data, white regions are pure CS$_2$.}
\label{fig:mean_event_rate-alpha_events} 
\end{center}
\end{figure}

\begin{figure}[htb]
\begin{center}
\includegraphics[width=\timePlotWidth\textwidth]{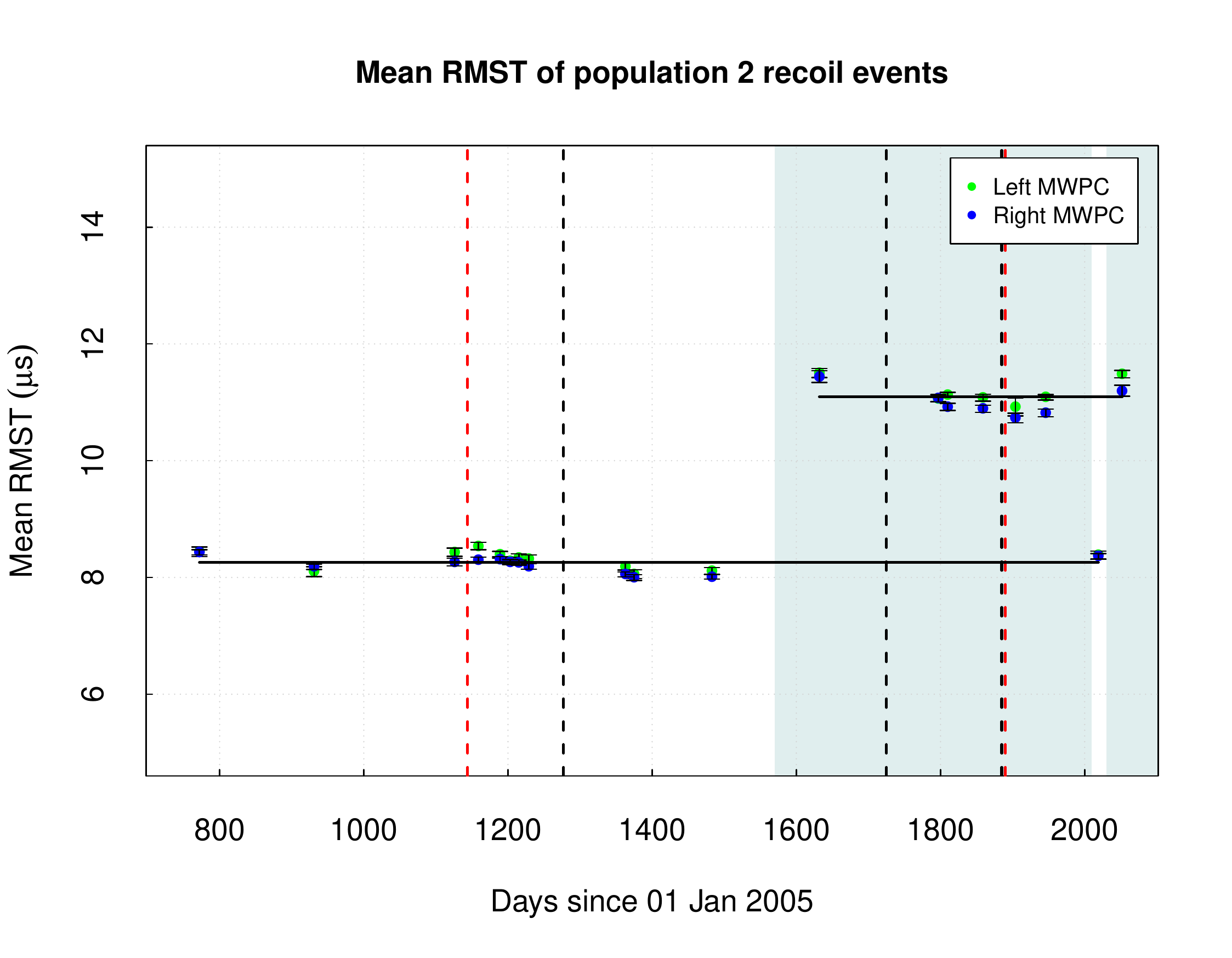}
\caption{RMST of population 2 events with mean rate during 40~Torr pure CS$_2$ data of 8.3$\pm$0.1\,\us~and during 30-10 Torr CS$_2$-CF$_4$ runs, at the same gas flow rate, of 11.1$\pm$0.1\,\us. Vertical dashed lines indicate actions described in section 3. Blue regions are CS$_2$-CF$_4$ data, white regions are pure CS$_2$.}
\label{fig:mean_rmst-mwpc_events} 
\end{center}
\end{figure}

The focus of this work is to detail changes in the four identified backgrounds as a function of time in a series of DRIFT-II runs over 5.5~years in which various operational changes were made.  Operations started in 2005, with subsequent upgrades to improve performance as follows: 
\begin{enumerate}
\item 9 Feb. 2008 (day 1134, first red dashed line in Figs. ~\ref{fig:mean_event_rate-cathode_events}--\ref{fig:mean_rmst-mwpc_events}) the central cathode wire plane was etched  to remove radioactive contaminants, such as  $^{210}$Pb, using a procedure similar to that of EXO-200 ~\cite{leonard2008}.  The 1\,m$^2$ cathode was placed in a bath of 3 molar nitric acid for two hours then washed in deionised water for 48 hours. The cathode was then outgassed under vacuum for 14 days.
\item 1 July 2008 (day 1277, first black dashed line in Figs. ~\ref{fig:mean_event_rate-cathode_events}--\ref{fig:mean_rmst-mwpc_events}) the MWPCs were etched, less aggressively, using one hour in 0.5 molar nitric acid, one hour deionised water, one hour 0.5 molar acid, 1 hour in deionised water, and 48 hours in fresh deionised water.
\item 22 Sept. 2009 (day 1725, second black dashed line in ~\ref{fig:mean_event_rate-cathode_events}--\ref{fig:mean_rmst-mwpc_events}) the right MWPC was replaced by an un-etched one.
\item 1 March 2010 (day 1885, third black dashed line in Figs. ~\ref{fig:mean_event_rate-cathode_events}--\ref{fig:mean_rmst-mwpc_events}) the right MWPC was changed back to the previously etched one.
\item 6 March 2010 (day 1890, second red dashed line in Figs. ~\ref{fig:mean_event_rate-cathode_events}--\ref{fig:mean_rmst-mwpc_events}) the wire plane cathode was replaced by a 0.9\,\um~Al-coated Mylar film.
\item For some runs the gas mixture was changed from pure CS$_2$ to CS$_2$-CF$_4$ (in Figs.~\ref{fig:mean_event_rate-cathode_events}--\ref{fig:mean_rmst-mwpc_events} shaded regions are for CS$_2$-CF$_4$). 
\end{enumerate}

The thin film cathode of (5) was developed to tackle the recoil-like backgrounds discussed in section 2, and is described in detail in Loomba et al. (2012) ~\cite{loomba2012}.  Briefly, making the cathode thin and thus more transparent to alpha particles increases the probability of background recoils being vetoed by detection of the associated alpha particle. Monte Carlo simulations predict that moving from a $20$~\si{\micro \metre} diameter steel wire cathode to a $0.9$~\si{\micro \metre} thin film alumnised Mylar cathode reduces the fraction of lost alpha particles from the decay of $^{218}$Po ($^{214}$Po) from 36\% (25\%) to 1.2\% (0.8\%). The difference is a consequence of the differing energies of the two alpha particles: 6~MeV and 7.69~MeV, respectively.

The remainder of this paper presents results of analysis of runs that are interspersed with the changes outlined above, using the reduced set of cuts described in section~\ref{sec:event_selection}. Rates of population 2 and 3 events were calculated as the areas of the two components of a double Gaussian fit to a histogram of RMST for events passing all cuts including the NIPs > 700 cut mentioned in section~\ref{sec:event_selection}. This is the y-projection, above 700~NIPs, of the plot appearing on the right hand side of figure~\ref{fig:moving_bg_plots_example}.

The mean rate of GPCCs over the period was found to be 54$\pm$3 events per day. A significant deviation from this occurred at day 931, with a rate of 20$\pm$3. This run had a nine-times faster flow rate, with the effect of flushing out Rn from the gas before it has time to decay, resulting in a lower GPCC count. The lack of a significant reduction in the corresponding recoil-like backgrounds suggests that these background events are dominated by long-lived $^{210}$Pb contamination already on the surface of the detector wires, and not the prompt RPRs.

We show in figure~\ref{fig:mean_event_rate-cathode_events} population 3 rates from recoil-like events over >1200 days. Three clearly delineated regions are seen, separated by the actions at days 1144 and 1890, as marked by red vertical dashed lines.  At the first action, in which nitric acid was used to etch the central cathode wires, the rate is reduced by a factor of $\sim 2$. At the second action, replacement of the cathode by a thin film version, the rate is again dramatically reduced. The actions on the MWPC planes (black dashed lines) have no significant effects on the rate of population 3. 

These results confirm that population 3 events indeed originate on the central cathode, and also demonstrate the success of the thin film cathode concept and effectiveness of dilute nitric acid for reduction of radioactive contamination. The factor $\sim 10$ rate reduction with the film is less than the factor $\sim 30$ reduction in the untagged RPR rate predicted by Monte Carlo simulations due in part to the overlap of populations 3 and 4, and in part to a new source of background introduced by the thin-film cathode. A detailed analysis to quantify the levels of U and Th contamination in the thin film will be the subject of a forthcoming paper.

Figure~\ref{fig:mean_event_rate-mwpc_events} shows the rate of population 2 recoil-like events.  Here, in contrast to population 3, the rate is different for the left and right MWPC. Recall that the red dashed vertical lines at days 1134 and 1890 indicate times of the cathode etch and switch to the thin film cathode, respectively. The black lines at days 1277, 1725 and 1885 correspond respectively to the etching of both MWPCs in nitric acid, the substitution of the right-hand side MWPC, and the subsequent replacement of the original right-hand side MWPC. No significant change in the rate of population 2 events resulted from the central cathode etch, in contrast to the large drop in rate in both detectors resulting from the etching of the MWPCs. This strongly supports the hypothesis that the population 2 events originate from contamination of the MWPCs, and again demonstrates the effectiveness of nitric acid etching. However, it is important to note that the measured rates have contributions from both population 4 and population 2 events, therefore we stop short of making a quantitative statement about the absolute rate of RPRs from the MWPCs. This will be addressed in a forthcoming paper.

On day 1725, the right MWPC was replaced by an older, unetched prototype. A large rate increase is seen in the population 2 region on the right MWPC, with no change on the left  and no increase in population 3. When the original MWPC was replaced on day 1885, the rate reduced again. This is further clear evidence that population 2 is indeed from the MWPCs.  The right detector also has consistently higher population 2 rate than the left detector. The levels of $^{210}$Pb contamination during manufacture of the wires varies from batch-to-batch, and it is therefore unsurprising to see two similar MWPCs with different background contamination levels.

Figure~\ref{fig:mean_event_rate-alpha_events} shows the rate of 1-MWPC alphas comprising events fully contained within one side of the detector. The asymmetry between left and right immediately suggests that the MWPCs are also the dominant source of the 1-MWPC events. This is supported by observation that the etching of the central cathode (day 1134) had no significant effect.  A further test relevant to this was performed from day 931, in which nine times the normal gas flow was used, drastically reducing any gas contamination. It can be seen that the 1-MWPC rate during this run is comparable with the mean rate after the MWPC etch (day 1277).  A particularly large increase in the 1-MWPC rate is observed on the right between days 1725 and 1885 when the replacement MWPC was in use, which is comparable to the increase seen in the event rate of the population~2 region during the same time period. Once again, possible population 4 contamination precludes the comparison of absolute rates. However, taken together, these large, coupled increases strongly suggests that these two backgrounds are closely related, consistent with the dominant source of both populations being the decay of radionuclides on the MWPC surfaces. When the alpha is oriented into the detector volume, a 1-MWPC event is produced, and when oriented into the MWPC wire the alpha particle is lost and the recoiling daughter nucleus produces a population~2 event.

Finally, we turn to changes made to the gas mixture between pure CS$_2$ and CS$_2$-CF$_4$.  As expected, no significant effect is seen in the RPR or alpha rates using the same flow rate. However, we show in figure~\ref{fig:mean_rmst-mwpc_events} a plot of the mean RMST of population~2 events vs. time.
In the white regions, corresponding to pure CS$_2$ runs, the mean is 8.3$\pm$0.1\,\us. A significantly larger mean, of 11.1$\pm$0.1\,\us, is observed during runs with 30-10 CS$_2$-CF$_4$, indicated by the blue shaded regions. In contrast the change in gas is observed to have no significant effect on the RMST for population~3, found to remain at 16.1$\pm$0.2\,\us.  The shift in RMST of population~2 events only is evidence again that these events arise form the MWPCs.
	
\section{Conclusions}
\label{sec:conclusions}
For the first time a consistent analysis of rare backgrounds has been performed in a low pressure Time Projection Chamber built for dark matter searches, namely DRIFT, covering several years and many detector configurations. The results provide new insights into the nature and sources of backgrounds relevant to directional searches for WIMP dark matter. The dominant backgrounds are found to be from four distinct populations. Strong evidence is presented that one population results from recoils of radon decay daughters on the surface of the central cathode plane.  A different population is found from recoils of decay daughters on the surface of the MWPC readout planes. In addition, a WIMP analysis of $^{55}$Fe x-ray calibration data has shown that a further low-energy population of recoil-like events are likely due to electron recoils following gamma-ray interactions.

Strategies to mitigate the backgrounds from alpha decay products have been successfully devised. Firstly, replacement of the wire plane central cathode with a thin film plane has substantially reduced the rate of recoils in the first population, and a forthcoming paper will present a detailed analysis of this improvement. This is a major advance for rare event search TPCS, including most directional dark matter detectors. A new acid etching technique was developed and applied to the MWPCs, which was found to reduce the second class of recoil event: those resulting from long-lived $^{210}$Pb deposits. Building on these results, a new texturised thin film central cathode is under development for DRIFT, with the goal of further reducing the RPR background revealed here, by ensuring that no straight path longer than $20$~\si{\micro \metre} exists inside the cathode~\cite{loomba2012}. Finally, a recent paper by Dan Snowden-Ifft has demonstrated that the addition of a small admixture of oxygen to the gas mixture provides a means by which to fiducialise events in the z-dimension, which has the potential to revolutionise the experiment's background-rejection capability \cite{Snowden2014}.

\acknowledgments
We thank Cleveland Potash Ltd. for continued support of the Boulby Laboratory.  M. Pipe thanks STFC for support on a PhD. grant.  The collaboration thanks the NSF for continued support of the DRIFT programme under grant numbers 1103420 and 1103511.

\end{document}